*Баліка С. Д.*

*Одеський національний університет імені І. І. Мечникова,
65026, вул. Дворянська, 2, Одеса, Україна
E-mail: svitlana.balika@onu.edu.ua*

## Визначення $\zeta$-потенціалу нанофлюїдів на базі розчинів електролітів за результатами вимірювань методами електричної спектроскопії та лазерної кореляційної спектроскопії

*У роботі обговорюється проблема вимірювання $\zeta$-потенціалу для нанофлюїдів на базі розчинів електролітів. Представлено теорію для дослідження впливу міжфазного (застійного) шару, зокрема, через такі параметри як $\zeta$-потенціал та товщина шару $u^*$ на ефективну провідність суспензій $\sigma_{eff}$. Теорія базується на методі компактних груп неоднорідностей, застосованому до систем провідних частинок з морфологією тверде ядро-проникна оболонка. Для запропонованої моделі встановлено аналітичну залежність провідності від цих параметрів. Із швидкості зміни провідності від концентрації для розбавлених нанофлюїдів залежність між $u^*$ і $\zeta$-потенціалом встановлюється однозначно. Тому для знаходження $\zeta$-потенціалу необхідно додатково визначити гідродинамічний радіус частинки, наприклад, через коефіцієнт дифузії Смолуховського-Ейнштейна. Таким чином, для знаходження $\zeta$-потенціалу пропонується провести серію незалежних одночасних вимірювань методами електричної спектроскопії та лазерної кореляційної спектроскопії.*

***Ключові слова:*** *$\zeta$-потенціал, електрична провідність, подвійний електричний шар, застійний шар*

**Вступ**

Електричні властивості зарядженої поверхні визначають величину поверхневого заряду та розподіл заряду в подвійному електричному шарі (ПЕШ) навколо неї [1-3]. Він складніший, ніж два шари, і деякі автори пропонують термін «електричний міжфазний шар». Електрокінетичний потенціал або $\zeta$-потенціал є важливим параметром заряджених поверхонь, що характеризує електричний стан міжфазної області. $\zeta$-потенціал визначається природою поверхні, її зарядом, концентрацією електроліту в розчині, природою електроліту та розчинника. Теоретично для однакових поверхонь із заданими параметрами $\zeta$-потенціал є чітко визначеною величиною. Однак, різні дослідники отримують різні $\zeta$-потенціали для однакових поверхонь, що вказує на чутливість потенціалу до різних параметрів, наприклад, незначної кількості домішок у розчині тощо. Серйозною проблемою є той факт, що для знаходження його числового значення використовуються різні модельні припущення, які базуються на певних уявленнях про структуру ПЕШ, лінеаризації нелінійних рівнянь, які описують електрокінетичні явища, і різні модельні наближення при аналізі. Його правильне визначення за допомогою електрокінетичних вимірювань вимагає чіткого розуміння меж застосовності теорії. Інтерпретація їх результатів і правильне використання цих теорій в межах їх застосування суттєво впливають на оцінку отриманих $\zeta$-потенціалів. У цьому плані питання про застосовність стандартної макроскопічної теорії до опису наночастинок є відкритим.

Слід також зазначити, що і саме питання про структуру ПЕШ є дискусійним. В багатьох стандартних моделях вважається, що ПЕШ складається з тонкого шару адсорбованих іонів (шар Штерна), що прилягають до поверхні частинки, та дифузного шару, у якому є надлишок іонів протилежного знаку і який підкоряється розподілу Больцмана. Електрокінетичні теорії якраз і використовують такі моделі. З іншого боку, сам дифузний шар насправді може складатися з двох частин, одна із яких є нерухомою, але може бути провідною – так званий застійний шар (*анг. stagnant layer*). Можна припустити, що структура цієї області складніша і насправді між шаром Штерна і дифузним шаром є навіть деяка перехідна область, існування якої зазвичай ігнорується стандартними теоріями електрофорезу. Це складна гідродинамічна задача.

В нашій роботі розглядається модель ПЕШ, яка врахує застійний шар та аналізується його вплив на ефективну електричну провідність нанофлюїду з розчином електроліту у якості базової рідини. Показано, що вона залежить від параметрів застійного шару, в першу чергу від його товщини. Встановлено, що швидкість зміни провідності розбавлених суспензій з концентрацією частинок дозволяє знайти функціональну залежність між товщиною цього шару та $\zeta$-потенціалом. Це означає, що маючи товщину із незалежних вимірювань, за результатами вимірювань провідності можна оцінити $\zeta$-потенціал в системі. Вкажемо, що для нанофлюїдів таким незалежним методом є метод кореляційної спектроскопії.

### 1. Моделі подвійного електричного шару

Згідно з найпростішою моделлю, ПЕШ складається з шару адсорбованих на поверхні іонів, які несуть певний електричний заряд, навколо якого формується дифузний електричний шар, у якому є надлишок іонів протилежного знаку. Отже, частинка має заряд прив'язаний до поверхні, та оточена зарядами в дифузному шарі, які є рухливими і підкоряються статистиці Больцмана. У деяких випадках вважають, що навіть поділу ПЕШ на шар Штерна та дифузний шар недостатньо для інтерпретації експериментів [4-6]. Шар Штерна поділяють на внутрішній шар Гельмгольца (ВШГ), обмежений поверхнею та внутрішньою площиною Гельмгольца (ВПГ), і зовнішній шар Гельмгольца (ЗШГ), розташований між ВПГ та зовнішньою площиною Гельмгольца (ЗПГ). Приклад такої моделі зображено на Рис.1 [7]. Такий поділ використовується, коли одні типи іонів специфічним чином адсорбуються на поверхні частинок, інші взаємодіють з поверхневим зарядом лише електростатичні сили. Оскільки виміряти поверхневий потенціал ізольованих частинок неможливо, необхідні додаткові модельні припущення. Тому найчастіше у дисперсних системах спочатку оцінюється поверхневий заряд, який залежить від властивостей компонентів та кислотності розчину [8], хімічної обробки частинок [9].

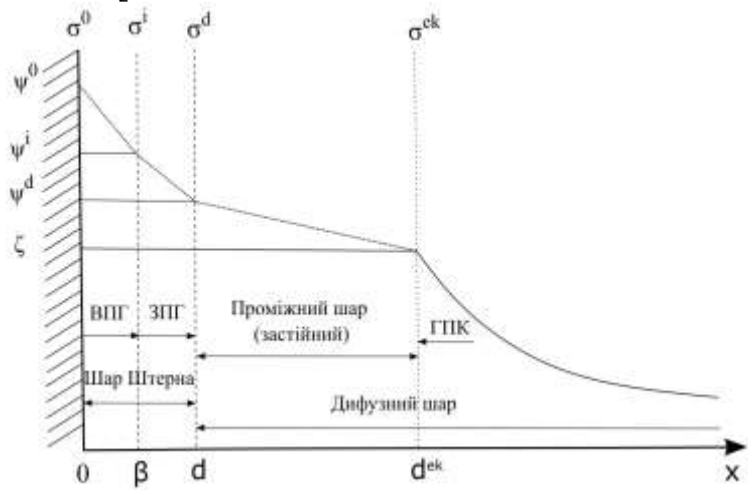

**Рис.1.** Схематичне представлення зарядів і потенціалів поблизу позитивно зарядженої поверхні. Область між поверхнею (електричний потенціал $\psi^0$, густина заряду $\sigma^0$) та ВПГ (відстань $\beta$ від поверхні) є незарядженою. ВПГ (електричний потенціал $\psi^i$; густина заряду $\sigma^i$) є областю спеціально адсорбованих іонів. Дифузійний шар починається при $x = d$ (ЗПГ), з потенціалом $\psi^d$ і густиною заряду $\sigma^d$. Площина ковзання або площина гідродинамічного зсуву розташована при $x = d^{ek}$. Потенціал у площині ковзання – це електрокінетичний або $\zeta$-потенціал; густина електрокінетичного заряду $\sigma^{ek}$. У цілому ця система електронейтральна: $\sigma^0 + \sigma^i + \sigma^d = 0$.

Молекулярно-динамічні розрахунки [10] показують, що при тангенціальному русі рідини вздовж зарядженої твердої поверхні, викликане зовнішнім електричним полем (електрофорез, електроосмосом), зазвичай дуже тонкий шар рідини прилипає до поверхні. Це гідродинамічно застійний (stagnant) шар, який тягнеться від поверхні на певну відстань $x < d^{ek}$, де знаходиться гідродинамічна площина ковзання (ГПК), цей шар гідродинамічно нерухомий, але може бути електропровідним. Фактично частинка разом із застійним шаром ведуть себе як одне ціле, радіус такої об'єднаної частинки часто називають фізичним радіусом частинки. ЗПГ інтерпретується як межа між недифузною та дифузною частинами ПЕШ, але точно локалізувати її складно. ГПК інтерпретується як межа між гідродинамічно рухомою і нерухомою рідиною. Насправді жодна із цих меж не є різкою. Однак рух рідини може бути утруднений в області сильної взаємодії іонів з поверхнею, тому іммобілізація рідини поширюється далі, ніж початок дифузної частини ПЕШ. Це означає, що на практиці $\zeta$-потенціал дорівнює або менше за величиною, ніж потенціал дифузного шару, $\psi^d$.

Як формується застійний шар та які його властивості досі не зрозуміло. Були припущення, що такий шар утворюється тільки на шорстких поверхнях [11] або він з'являється внаслідок в'язкоелектричного ефекту, тобто збільшується в'язкість приплилої рідини, що викликано поляризацією подвійного шару [12]. Однак, застійні шари можуть утворюватися і на однорідних гладких поверхнях, і на незаряджених поверхнях. Стандартні теорії таких електрокінетичних явищ передбачає розв'язок рівнянь руху з рівняннями поля та врахування певних граничних умов [13-15]. У цих теоріях частинка разом із застійним шаром розглядається як тверде інертне тіло. Поверхнева провідність у двошаровій частині за площиною ковзання враховується автоматично. Було виявлено, що для застійних шарів, що примикають до твердих неворсистих поверхонь, ці рухливості трохи нижчі, ніж в основній матриці.

Теорія Смолуховського [2] справедлива для непровідних частинки за умови, що радіус кривизни $a$ значно перевищує радіус Дебая $\kappa^{-1}$: $\kappa a >> 1$. При цьому, впливом поверхневої провідності як у дифузній, так і у внутрішній частинах ПЕШ та його поляризацією, нехтують. Для достатньо високих потенціалів теорія непридатна. У випадку концентрованих систем не можна нехтувати можливістю перекриття подвійних шарів сусідніх частинок.

У зв'язку з переліченим, оцінки $\zeta$-потенціалу, особливо для систем наночастинок, видається актуальним. На відміну від існуючих моделей, ми включаємо у свою модель перехідну область між шаром Штерна та ГПК, яка є початковою частиною дифузного ПЕШу Будуємо теорію електропровідності суспензії таких частинок з урахуванням такої внутрішньої структури і виводимо для електролітів (суспензій) вирази для ефективної провідності системи.

В нашій теорії ми надмірно не деталізуємо властивості застійного шару ПЕШ, основними параметрами виступають його товщина і відносна величина провідності. Модель значно поглиблює недавні теоретичні результати [16,17], які виявилися дуже ефективними при застосуванні до невпорядкованих систем твердих частинок, твердих композитних електролітів і полімерних композитних електролітів. Основна особливість теорії полягає в тому, що мезоструктура і властивості суспензії описуються за допомогою моделі макроскопічно однорідною, ізотропною дисперсією частинок з морфологією тверде ядро–проникна оболонка, вкраплених в однорідну матрицю з провідністю $\sigma_0$. Оболонки частинок електрично неоднорідні в радіальному напрямку, профіль їх провідності $\sigma_2(u)$ неперервний. Компоненти, що перекриваються, підкоряються правилу домінування, згідно з яким локальне значення провідності визначається відстанню від даної точки до найближчої частинки.

Електродинамічна гомогенізація моделі враховує ефекти поляризації та кореляції багатьох частинок і є внутрішньо замкнутою.

## 2. Модель провідності нанофлюїдів

Сформулюємо основні ідеї роботи. Ми побудували теорію, яка дозволяє оцінити $\zeta$-потенціал через ефективну провідність $\sigma_{eff}$. Маємо формулу для провідності, при певних припущеннях щодо провідності застійного шару, отримаємо ефективну провідність в залежності від товщини застійного шару і $\zeta$-потенціалу. Оскільки к$a$ і концентрація частинок величини контрольовані можемо знайти швидкість зміни провідності з концентрацією, яка експериментально вимірюється. Отримаємо замкнений вираз між товщиною шару і $\zeta$-потенціалом. Якщо знаємо товщину можемо оцінити потенціал, або навпаки, знаючи потенціал, знайти товщину шару. Далі незалежним методом можна знайти гідродинамічний радіус частинки, наприклад, методами лазерної кореляційної спектроскопії, коли розглядається розсіяння світла в нанофлюїдах. Знаючи коефіцієнт дифузії Смолуховського-Ейнштейна, можемо знайти фізичний радіус частинки. Такі дослідження проводились для нанофлюїдів ізопропанол/$Al_2O_3$ [18]. Знаючи радіус і знаючи к$a$ можемо знайти $\zeta$-потенціал. Цей метод не вимагає ніяких додаткових припущень. Достатньо провести в таких системах серію незалежних, але одночасних вимірів, знайти провідність та радіус. Маючи дані наночастинок, можна знайти товщину області, яка прилипає до частинки, та оцінити $\zeta$-потенціал.

Теорія базується на результатах, отриманих методом компактних груп [19-21] для систем частинок тверде ядро–проникна оболонка [16,17]. Суспензія моделюється як суміш твердих ядер (які відповідають частинкам реальних систем) радіусом $a$, провідністю $\sigma_1$ та об'ємною концентрацією $c$, і прилеглих до них проникних оболонок, що складаються з двох частин: внутрішньої оболонки з відносними товщиною $u^* = (R-a)/a$ і зовнішньої оболонки (пов'язаної з рухливою частиною дифузного ПЕШ). $R$ – радіус частинок разом із прилеглим застійним шаром. Оболонки є електрично неоднорідними, з розподілом електропровідності $\sigma_2(u)$. Частинки і базова рідина з провідністю $\sigma_0$ вкраплені в матрицю з провідністю $\sigma_{eff}$. Для оцінки об'ємної концентрації ядер разом з оболонками $\varphi(c,u)$ використовуємо дві граничні моделі – вільно проникних [22,23] та твердих оболонок.

У рамках методу компактних груп суспензія розглядається як сукупність макроскопічних областей (компактних груп) з лінійними розмірами, набагато меншими за довжину хвилі зовнішнього поля в системі. З одного боку, такі групи мають симетрію та макроскопічні властивості всієї системи. З другого боку, у порівнянні з довжиною хвилі поля вони фактично є точковими. Використовуючи методи теорії узагальнених функцій і спеціальне представлення пропагатора електромагнітного поля, у довгохвильовому наближенні можна виокремити внески цих груп, уникаючи при цьому деталізації процесів перевипромінювання та кореляцій між частинками. Доводиться, що ці внески являються визначальними та формують ефективні квазістатичні діелектричні та провідні властивості суспензії.

Головне рівняння для ефективної провідності з високопровідним внутрішнім шаром $\sigma_2(u) \gg \sigma_{\text{eff}}$ для $u \in (0, u^*)$ та непровідними частинками $\sigma_1 \to 0$ має вигляд [24]:

$$[1 - \varphi(c, u^*)]\frac{\sigma_0 - \sigma_{\text{eff}}}{2\sigma_{\text{eff}} + \sigma_0} + \varphi(c, u^*) - \frac{3}{2}c - $$
$$3\sigma_{\text{eff}} \int_{u^*}^{\infty} \frac{\partial \varphi(c, u)}{\partial u}\left[\frac{1}{2\sigma_{\text{eff}} + \sigma_2(u)} - \frac{1}{2\sigma_{\text{eff}} + \sigma_0}\right]du = 0 \quad (1)$$

Схожі співвідношення виводимо і для випадків $\sigma_2(u) \ll \sigma_{\text{eff}}$ та $\sigma_2(u) \approx \sigma_{\text{eff}}$. Усі доданки цього рівняння взаємопов'язані та мають прозорий фізичний зміст: перший доданок описує внесок базової рідини; другий – це внесок частинок, ефективна провідність яких включає ефект іонного транспорту в шарі Штерна; третій описує внесок внутрішньої частини дифузного шару; четвертий – внесок зовнішньої частини дифузного шару. За площиною ковзання в області $u > u^*$ провідність формується рухливими іонами в дифузному шарі та підкоряються розподілу Больцмана. Профіль електричної провідності в дифузній частині ПЕШ для $u > u^*$:

$$\sigma_2(u) = e\mu_+(u)Z_+ n_+^{(0)} e^{-Z_+ y(u)} - e\mu_-(u)Z_- n_-^{(0)} e^{-Z_- y(u)} \quad (2)$$

де $e$ – елементарний заряд; $\mu_+, \mu_-$ - рухливості відповідно катіонів та аніонів із зарядовими числами $Z_+, Z_-$ та середніми концентраціями $n_+^{(0)}, n_-^{(0)}$; $y(u)$ – розподіл потенціалу в дифузній частині ПЕШ. Знаходження останнього є нетривіальною багаточастинковою задачею, але у випадку, коли перекриттям оболонок можна нехтувати, використовуємо прості моделі, наприклад, Гуї-Чепмена для плоскої моделі:

$$y(u) = 2\ln\left[\frac{1 - |\gamma| e^{-\kappa a u}}{1 + |\gamma| e^{-\kappa a u}}\right], \quad \gamma \equiv \tanh\left(\frac{e\zeta}{4kT}\right) \quad (3)$$

де $k$ – стала Больцмана; $T$ – температура.

Для розбавлених суспензій знаходимо величину $\frac{1}{\sigma_0}\frac{\partial \sigma_{\text{eff}}}{\partial c}$, яку експериментально можна поміряти:

$$\frac{1}{\sigma_0}\frac{\partial \sigma_{\text{eff}}}{\partial c} = 3(1 + u^*)^3 - \frac{9}{2} + 9\int_{u^*}^{\infty}(1 + u)^2 \frac{\sigma_2(u) - \sigma_0}{2\sigma_0 + \sigma_2(u)}du \quad (4)$$

Скористаємося формулою (4) для оцінки знизу товщини застійного шару. Згідно з експериментальними даними [25], для суспензії частинок латексу B радіусом 235 нм у водному розчині HCl молярністю $M = 0.005$ ($\kappa a - 54.6$) при значенні $e\zeta/kT = -3.22$ похідна $\frac{1}{\sigma_0}\frac{\partial\sigma_{eff}}{\partial c} = -0.85 \pm 0.07$. Для цих даних формула (4) дає $u^* = 0.052$, що відповідає товщині застійного шару приблизно 12 нм. Можна очікувати, що для наночастинок з паспортним радіусом $\leq 100$ нм застійний шар з такою самою товщиною повинен збільшувати їх фізичний розмір на $\geq 12\%$. Таке зростання розміру частинок можна зареєструвати оптичними методами.

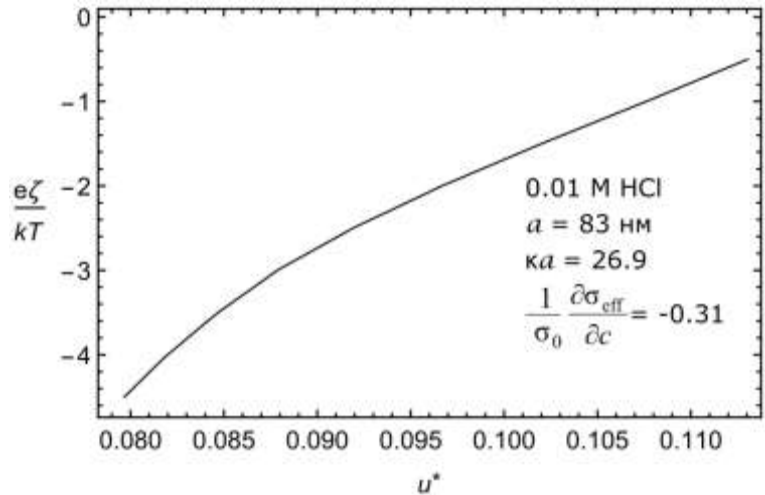

**Рис.2.** Приклад залежності $\zeta$- потенціалу від $u^*$ для суспензій латексних частинок радіусом $a = 83$ нм $(\kappa a = 26.9)$ у водному розчині 0.01 М HCl для експериментального значення $\frac{1}{\sigma_0}\frac{\partial\sigma_{eff}}{\partial c} = -0.31$ [25].

У роботі [18] методами лазерної кореляційної спектроскопії [26,27] визначалися характерні розміри оптичних неоднорідностей нанофлюїди ізопропанол/$Al_2O_3$. Вимірювання проводилися на оригінальній установці методом гомодинування [28]. Характерні фізичні розміри наночастинок визначалися за допомогою результатів вимірювання кореляційної функції інтенсивність-інтенсивність розсіяного у гауссовому наближенні для флуктуацій:

$$G^2(\tau) = \langle I(0)I(\tau)\rangle = A + B\exp(-2Dq^2\tau) \qquad (5)$$

де $D = \frac{kT}{6\pi\eta r}$ – коефіцієнт дифузії Смолуховського-Ейнштейна, $q$ – зміна хвильового вектора внаслідок розсіяння.

Характерний розмір частинок визначався як гідродинамічний радіус [27]

$$r = \frac{8\pi n^2 \sin^2\left(\frac{\theta}{2}\right)kT\tau_c}{3\eta\lambda_0^2} \qquad (6)$$

де $n$ – показник заломлення розчину, $\theta$ – кут розсіяння, $\tau_c$ – час кореляції, $\eta$ – коефіцієнт динамічної в'язкості, $\lambda_0$ – довжина хвилі падаючого випромінювання.

Дані у роботі [18] свідчать про зміну гідродинамічного радіуса частинок $Al_2O_3$ від 53 до 86 (тобто до 62%) нм при збільшенні масової концентрації наночастинок в ізопропанолі від 0,036 до 4,2%.

**Висновок**

На основі проведеного аналізу показано, що визначення $\zeta$- потенціалу нанофлюїдів можливе шляхом комбінації методів електричної спектроскопії та лазерної кореляційної спектроскопії. Запропонована теорія дозволяє уникати неоднозначності при інтерпретації результатів електрофоретичних вимірювань на основі класичних моделей. У запропонованій моделі ефективна провідність виражена через товщину застійного шару і $\zeta$-потенціал. Оцінюючи швидкість зміни провідності в залежності від концентрації методом електричної спектроскопії, отримано залежність між товщиною застійного шару $u^*$ та $\zeta$-потенціалом. Аналізуючи розсіяння світла незалежним методом лазерної кореляційної спектроскопії, можна знайти гідродинамічний радіус частинки. Маючи паспортні дані наночастинок, далі можна оцінити $u^*$ та знайти $\zeta$-потенціал.

Представляють інтерес організація й проведення відповідних експериментів та практична апробація запропонованої методики.

**Література**

***Balika S.D.***

**Determination of $\zeta$-potential of nanofluids based on electrolyte solutions from the measurements by the methods of electrical spectroscopy and laser correlation spectroscopy**

*The work discusses the problem of measurement of the $\zeta$-potential for electrolyte-based suspensions of nanoparticle. A theory is presented for the effect of the diffuse electric double layer, including the interphase (stagnant) layer, on the*

*effective conductivity of such suspensions. The theory is based on the method of compact groups of inhomogeneities applied to a model system of hard-core–penetrable-shell particles embedded together with the base liquid in a uniform host of the conductivity* $\sigma_{\text{eff}}$. *The cores represent the particles. The shells are electrically inhomogeneous in the radial direction, their conductivity profile being a continuous function* $\sigma_2(r)$. *The overlapping components obey the rule of dominance, according to which the local conductivity value is determined by the distance from the point to the center of to the nearest particle. This model is possible to analyze rigorously in the static limit. The desired conductivity is found from the integral relation which allows us to express the electrical conductivity in the terms of the* $\zeta$*-potential, thickness of the stagnant layer, matrix molarity, etc. The theory reveals the existence of different scenarios of behavior of the conductivity depending on the geometrical and electrical parameters of the stagnant layer. It is shown that the functional dependence between the thickness and* $\zeta$*-potential can be obtained from the rate of change of the conductivity with concentration for diluted suspensions; this rate can be measured experimentally. It is pointed out that the hydrodynamic radius of a nanoparticle can be obtained from the Smoluchowski-Einstein diffusion coefficient, which can be measured by the method of laser correlation spectroscopy. We give all necessary estimates and relations to demonstrate the opportunity of measuring the* $\zeta$*-potential by analyzing together the results obtained by the indicated methods. Namely, in order to find the* $\zeta$*-potential, it is necessary to simultaneously do a series of independent measurements: find the slope of the conductivity and estimate the thickness of the stagnant layer.*

**Keywords:** $\zeta$-potential, electric conductivity, electric double layer, stagnant later